\documentclass[12pt]{article}
\usepackage{epsfig}
\usepackage{amssymb}
\usepackage{bbm}
\usepackage{graphicx}

\newcommand{\be}{\begin{equation}}
\newcommand{\ee}{\end{equation}}
\newcommand{\beq}{\begin{equation}}
\newcommand{\eeq}{\end{equation}}
\newcommand{\bea}{\begin{eqnarray}}
\newcommand{\eea}{\end{eqnarray}}
\newcommand{\br}{\begin{eqnarray}}
\newcommand{\er}{\end{eqnarray}}
\newcommand{\ba}{\begin{array}}
\newcommand{\ea}{\end{array}}
\newcommand{\bi}{\begin{itemize}}
\newcommand{\ei}{\end{itemize}}
\newcommand{\bn}{\begin{enumerate}}
\newcommand{\en}{\end{enumerate}}
\newcommand{\bc}{\begin{center}}
\newcommand{\ec}{\end{center}}

\def\unity{{\hbox{1\kern-.8mm l}}}

\def\tv#1{\vrule height #1pt depth 5pt width 0pt}

\newcommand{\gsim}{\lower.7ex\hbox{$\;\stackrel{\textstyle>}{\sim}\;$}}
\newcommand{\lsim}{\lower.7ex\hbox{$\;\stackrel{\textstyle<}{\sim}\;$}}

\def\mysection#1{\noindent {\bf #1} }

\begin{document}

\begin{titlepage}
\begin{flushright}
CPHT-RR-060.1005 \\ LPT-ORSAY-05-70 \\ Saclay T05/167 \\ 
{\tt hep-ph/0511001}
\end{flushright}

\vskip.5cm
\begin{center}
{\huge \bf Classical running of neutrino masses \\
\vspace{.2cm}
from six dimensions}
\vskip.1cm
\end{center}
\vskip0.2cm

\begin{center}
{\bf
{E. Dudas}$^{a,b}$,
{C. Grojean}$^{c}$,
{S.K. Vempati}$^{a}$}
\end{center}
\vskip 8pt

\begin{center}
$^{a}$ {\it Centre de Physique Theorique\footnote{Unit{\'e} mixte du CNRS et de l'EP, UMR 7644.}, Ecole Polytechnique, 91128 Palaiseau Cedex, France} \\
\vspace*{0.1cm}
$^{b}$ {\it
 LPT\footnote{Unit{\'e} mixte du CNRS, UMR 8627.},
B{\^a}t. 210, Univ. de Paris-Sud, F-91405 Orsay, France} \\
\vspace*{0.1cm}
$^{c}$ {\it Service de Physique Th\'eorique\footnote{Unit{\'e} de recherche associ au CNRS, URA 2306.},
CEA Saclay, F91191 Gif--sur--Yvette, France}\\
\vspace*{0.3cm}
{\tt  Emilian.Dudas@cpht.polytechnique.fr,
grojean@spht.saclay.cea.fr, vempati@cpht.polytechnique.fr}
\end{center}

\vglue 0.3truecm

\begin{abstract}
\vskip 3pt
\noindent
We discuss a six dimensional mass generation for the neutrinos.
Active neutrinos live on a three-brane and interact via a brane
localized mass term with a bulk six-dimensional standard model
singlet (sterile) Weyl fermion, the two dimensions being
transverse to the three-brane.
  We derive the physical neutrino mass spectrum and show that the
active neutrino mass and Kaluza--Klein masses have a
logarithmic cutoff divergence related to the zero-size limit of the
three-brane in the transverse space. This translates into a renormalisation
group running of the neutrino masses above the Kaluza--Klein
compactification scale coming from classical effects, without any new
non-singlet particles in the spectrum. For compact radii in the
eV--MeV range, relevant for neutrino physics, this scenario predicts
running neutrino masses which could affect, in particular,
neutrinoless double beta decay experiments.
\end{abstract}

\end{titlepage}

\newpage

%\renewcommand{\thefootnote}{(\arabic{footnote})}

%%%%%%%%%%%%%%%%
\section{Introduction and Conclusions}
%%%%%%%%%%%%%%%%

Extra dimensional models have become quite popular in the past few
years with the realisation that they could exist at far lower scales compared
to the Planck scale, $M_{pl}$, and thus they could have an impact at the
TeV scale physics~\cite{add1,ig,rs}. Some of the significant implications
could be for the gauge coupling unification~\cite{ddg1}, solutions
to the hierarchy problem~\cite{add1,rs} each having its own phenomenological
impact like for example, collider searches~\cite{Giudice:1998ck} and precision measurements of
standard electroweak parameters~\cite{Delgado:1999sv}.

The existence of the extra spacetime dimensions at some scale closer or
higher than the weak scale also gives much scope to build models related
to fermion masses and mixings, particularly for light neutrinos \cite{nmass},
qualitatively rather different from the traditional 4D seesaw
mechanism~\cite{seesaw}. In this case,
a singlet neutrino (right handed) could be allowed to propagate in the
bulk whereas the SM (left-handed) neutrino would be confined to the 3-brane,
leading to new possibilities of neutrino masses and mixing~\cite{ddg2,addm,nrs}.
While this possibility has been explored in detail both in model building
as well as in terms of phenomenological analysis~\cite{rr}, most of these
studies have
been confined to the case where there is only one additional dimension.
It is necessary to extend these analysis for higher number of extra
dimensions. This is because new features could arise in higher dimensional
field theories which could give rise to different phenomenology. In particular, interesting 6D models have been constructed  to break the electroweak symmetry through non-trivial Wilson lines~\cite{GaugeHiggs6D}, to guarantee the proton stability up to dimension fifteen operators~\cite{Appelquist:2001mj}, to predict the number of chiral generations~\cite{Dobrescu:2001ae}, to provide a dark matter candidate~\cite{KKDM}, to construct realistic GUT models~\cite{6DGUT} etc. 6D models have also been constructed to reproduce the neutrino mass spectrum~\cite{6Dneutrinos}.

In the present work, we will study the case of neutrino masses in
a six dimensional model. In six dimensional models, it has been
known for some time that orbifold compactification produces some
peculiar properties such as `tree level' renormalisation of
coupling constants~\cite{ggh,gw}, whereas one-loop quantum effects
in open string theories have a similar interpretation~\cite{ab}.
While this has been known for the scalar case for some time, we
will study this property in detail for the case of neutrinos with
a brane localized mass term on a $T^2/\mathbb{Z}_2$ orbifold. We find that
the physical neutrino mass (and also higher Kaluza--Klein modes)
have a logarithmic divergence related to the brane thickness. We
find this result by two different methods. First by
diagonalizing the (infinite) mass matrix in the KK basis, along
the lines of~\cite{ddg2,addm,Dobrescu:2004zi}. Second by considering the bulk
propagation of the fields with appropriate boundary conditions due to
the orbifold projection in presence of  brane-localized operators~\cite{BCs}.

We show that its interpretation is similar to the brane-localized
scalar mass term discussed in~\cite{gw} and it can be renormalised in
a similar manner by adding a neutrino Dirac mass brane
counterterm. As a result, the neutrino mass runs, more precisely
increases with energy, above the compactification scale $R^{-1}$,
where for simplicity we consider the case of two equal radii. For
radii in the eV to MeV range, the effect of this classical running
has no counterpart in four dimensions, since it arises without the
presence of any new particle charged under the Standard Model
gauge group. This can be tested in processes with off-shell
neutrinos, like the neutrinoless double beta decay, where the
increase in the neutrino mass at GeV energies enhances the
amplitude of the process.

The paper is organized as follows. Section 2 describes the
example of a six-dimensional scalar field with four-dimensional
(brane-localized) mass term in one of the orbifold fixed points,
in the simplest orbifold compactification.  We diagonalize
explicitly the mass matrix and show that the physical masses
(eigenvalues) have a logarithmic dependence on a UV cutoff, related
to the thickness of the brane where the mass term was inserted.
The same result is then obtained by solving bulk field equations
with appropriate boundary conditions at the mass distribution,
giving therefore a clear meaning of the ultraviolet cutoff in
terms of the profile of the mass distribution. In Section~3 an
off-shell analysis along the lines of~\cite{ggh,gw} suggests that
the cutoff dependence has a natural interpretation in terms of
renormalisation group running of the localized mass term, between
the compactification scale and the ultraviolet cutoff.
In Section~4 we pass to the case of interest in the present
paper, four-dimensional localized active neutrino mixing via a
localized Dirac mass term with a bulk six-dimensional singlet Weyl
fermion. We show that the eigenvalue equation in
this case is exactly the same as in the scalar field with
localized mass discussed in Section~2. We then break the lepton
number conservation by adding a brane-localized Majorana mass term for the bulk
field and work out again the physical masses and their
corresponding energy running. In Section~5 we discuss
higher-dimensional neutrino oscillations and in particular the
probability of regeneration of active neutrinos.
We end up in Section~6 with a quick
survey of the phenomenological consequences of this class of
models coming from the fundamental scales and the running, which
can be tested in particular in the neutrinoless double beta decay experiments.

%%%%%%%%%%%%%%%%%%%%%%%%%%%%%%%%%%%%%%%%%%%%%%%%%%%%%%%%%%%%%%%%%%%%%%%%5
\section{The Scalar Case}

We will discuss the 6D formalism in detail both in the scalar as
well as the fermion case. We consider a $T^2/\mathbb{Z}_2$ orbifolding in
the two dimensional compact space. As a starting point, let us set
the bulk mass to zero and add a localized mass term at the origin
of the 2D compact space. The corresponding action
reads\footnote{We are using a $(+,-,-,-,-,-)$ metric. The index $M$ denotes bulk coordinates and runs from $0,1,2,3,5,6$, while $\mu=0,1,2,3$ denotes brane coordinates. We'll use either $x^{5,6}$ or $y_{1,2}$ to denote the two extra dimensions. Finally, $\partial_{1,2}$ is a short-handed notation for $\partial_{y_{1,2}}$} :
\begin{equation}
    \label{scalaraction}
S =  \frac{1}{2} \int d^4x d^2 y \
\left( (\partial^M \Phi )(\partial_M \Phi) -
h_2 \ \Phi^2 \delta^2( {\bf y})
 \right) \ ,
\end{equation}
where $h_2 > 0 $ is a dimensionless
coupling in the natural 6D units for which the scalar field $\Phi$
has dimension two. The coupling $h_2$ is localized
at the origin of the compact space.  The field equation is free in the bulk and
has a delta function source at the origin
\begin{equation}
    \label{sc1}
\partial_M \partial^M \Phi + h_2  \Phi \, \delta^2 ({\bf y})= \ 0 \ .
\end{equation}
We will consider
compactification on the orbifold $T^2/\mathbb{Z}_2$, acting as the
reflection $(y_1,y_2) \rightarrow (-y_1,-y_2) $. This orbifold has
four fixed points as summarised below in Fig.(1). 
The fixed points and the corresponding
$\mathbb{Z}_2$ transformations of the coordinates around
those points are summarised as:
\begin{equation}
     \label{orbifoldfix}
\begin{array}{cccc}
y_1 \to -y_1 ~& y_1 \to -y_1 + 2 \pi R_1~ &~ y_1 \to -y_1 ~&~
y_1 \to -y_1 + 2 \pi R_1 \\
y_2 \to -y_2 ~&~ y_2 \to -y_2  ~&~ y_2 \to -y_2 + 2 \pi R_2
~&~ y_2 \to -y_2 + 2 \pi R_2 \\
(0,0)~&~ (\pi R_1,0) ~&~ (0, \pi R_2)~&~ (\pi R_1, \pi R_2).
\end{array}
\end{equation}
In complex notation, the action of $\mathbb{Z}_2$ on the compact space
is a  two-dimensional $\pi$ rotation,
$Z_2(y_1 +iy_2)$ = $e^{i \pi}(y_1 + i y_2)$.  Let us now proceed
to study the KK spectrum.

\begin{figure}[ht]
\centerline{\includegraphics[scale=0.6]{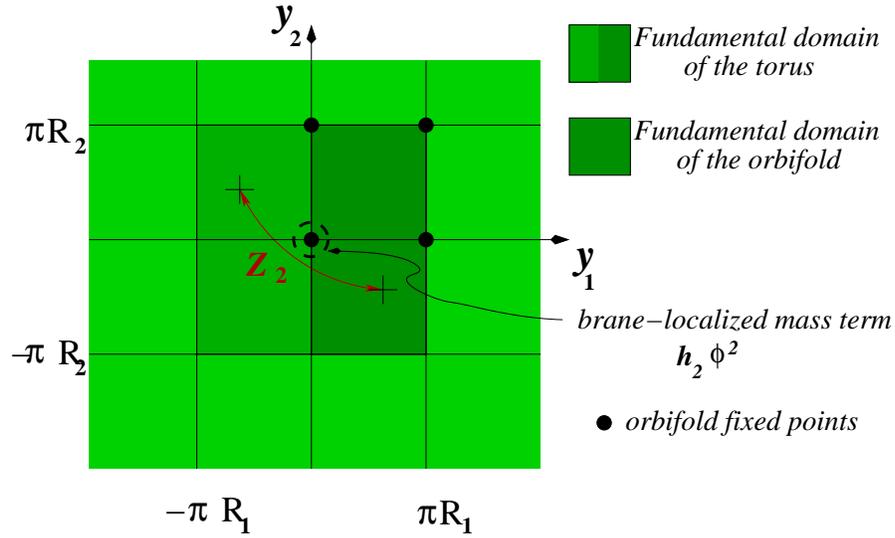}}
\caption{$\mathbb{Z}_2$ orbifold with a brane localized mass term for the scalar field at the origin.}
\end{figure}

%%%%%%%%%%%%%%%
\subsection{The KK approach}
%%%%%%%%%%%%%%%

If the scalar field $\Phi$ is even under the orbifold action,
it can be decomposed on a complete basis formed by the cosine functions:
 \begin{equation}
\Phi(x,{\bf y})
 =
 \sum_{(k_1,k_2) \in \mathcal{I}}
 \langle y_1, y_2 | k_1, k_2 \rangle
 \, \phi_{(k_1,k_2)} (x)
 \end{equation}
 with
\begin{equation}
\langle y_1, y_2 | k_1, k_2 \rangle
=
\frac{1}{\sqrt{2 \pi^2 R_1 R_2 }}
\frac{\cos \left( \frac{k_1}{R_1} y_1 +  \frac{k_2}{R_2} y_2 \right)}
{\sqrt{2^{\delta_{k_1 0}\, \delta_{k_2 0}}}}
\ ,
\end{equation}
The indices $k_{1,2}$ belong to the set $\mathcal{I}$
\begin{equation}
	\label{eq:setI}	
\mathcal{I} = 
\left\{ 
(0;0), (1\ldots \infty; 0), (0;1\ldots \infty),
(1\ldots \infty; 1\ldots \infty),
(1\ldots \infty; -\infty \ldots -1)
\right\}\ .
\end{equation}
The scalar action~(\ref{scalaraction}) then takes the following form
after integration over the two extra dimensions
\begin{equation}
    \label{eq:4Daction}
\mathcal{L} =
\mathcal{L}_{kin}
- \frac{1}{2}
\sum_{(k_1,k_2) \in \mathcal{I}}
\left( \frac{k_1^2}{R_1^2} + \frac{k_2^2}{R_2^2} \right)
\phi_{(k_1,k_2)}^2
-
\frac{\bar{m}^2}{2}
\left( \sum_{(k_1,k_2) \in \mathcal{I}}
\frac{\sqrt{2}}{\sqrt{2^{\delta_{k_1 0}\, \delta_{k_2 0}}}} \, \phi_{(k_1,k_2)}
\right)^2
\ ,
\end{equation}
where
\begin{equation}
\bar{m}^2 \equiv \frac{h_2}{4 \pi^2 R_1 R_2}
\end{equation}
 is the naive (volume
suppressed) four dimensional lightest scalar mass, which is typically of the
order and slightly smaller than the compactification mass scale.  The mass term of the 4D
action~(\ref{eq:4Daction}) is
\begin{equation}
\mathcal{L}_{\rm mass} =
- {1 \over 2}
\sum_{(k_1,k_2),(p_1,p_2) \in \mathcal{I}}
 \phi_{(k_1,k_2)} \,
\mathcal{M}^2_{(k_1,k_2),(p_1,p_2)} \,
 \phi_{(p_1,p_2)}
\end{equation}
with the mass matrix given by
\begin{equation}
    \label{massmatrix1}
\mathcal{M}^2_{(k_1,k_2),(p_1,p_2)}  =
\frac{2 \bar{m}^2}{\sqrt{2^{\delta_{k_1 0}} \, 2^{\delta_{k_2 0}}}}
+ \left( \frac{k_1^2}{R_1^2} + \frac{k_2^2}{R_2^2} \right) \delta_{k_1,p_1} \delta_{k_2,p_2} \ .
\end{equation}
The diagonalization of this  mass matrix will define the KK mass eigenstates.

%%%%%%
\subsection{Eigenvalue Analysis}
%%%%%%

Let us now try to find the eigenvalues and eigenvectors of the
mass matrix~(\ref{massmatrix1}). The characteristic equation
is given
by
\begin{equation}
    \label{charectersticeq}
 \mathcal{M}^2 \Psi_m = m^2 \Psi_m,
\ee
where $m^2$ represents the eigenvalues and $\Psi$ is the eigenvector
in the basis $\left| k_1,k_2 \right. \rangle_{(k_1,k_2) \in \mathcal{I}}$ defined in the previous section
($\Psi_{(k_1,k_2)} = \langle k_1, k_2 | \Psi_m \rangle$ ).
The matrix equation~(\ref{charectersticeq}) is equivalent to the infinite set of explicit equations for every
$(k_1,k_2) \in \mathcal{I}$
\begin{equation}
    \label{chaexp}
\sqrt{2} \bar{m}^2 \Psi' = \sqrt{2^{\delta_{k_1 0}\, \delta_{k_2 0}}}
\left( m^2 - \frac{k_1^2}{R_1^2} - \frac{k_2^2}{R_2^2} \right)
\Psi_{(k_1,k_2)},
\end{equation}
where we have defined
\begin{equation}
\Psi' =
\sqrt{2}
\sum_{(k_1,k_2) \in \mathcal{I}}
\frac{1}{\sqrt{2^{\delta_{k_1 0}\, \delta_{k_2 0}}} }
\Psi_{(k_1,k_2)}
 \ .
\end{equation}
The solution of the equations~(\ref{chaexp}) is simply given by
\begin{equation}
    \label{egvec}
\Psi_{(k_1,k_2)} =
\frac{1}{\sqrt{2^{\delta_{k_1 0}\, \delta_{k_2 0}}} }
\ \frac{\mathcal{N}}{m^2 - k_1^2/R_1^2- k_2^2/R_2^2}
 \ ,
\end{equation}
where $\mathcal{N}$ is a normalization constant independent of $(k_1,k_2)$.
Putting this solution back in the equation~(\ref{chaexp}) for $k_1=k_2=0$, we obtain the eigenvalue equation we were searching for:
\begin{equation}
    \label{eigenvalue}
\frac{1}{\bar{m}^2} =
\sum_{k_1, k_2 = -\infty}^\infty \frac{1}{m^2 - k_1^2/R_1^2 - k_2^2/R_2^2 }
\ .
\end{equation}
Note that, in this last equation,  the sums have been conveniently rewritten from $k_i=-\infty$ to 
$+\infty$~(see the identity~(\ref{eq:identitysum}) of the Appendix~A).

We want to find an estimate for the lightest solution, $m^2$, of the
eigenvalue equation. We are first going to evaluate the sums
in~(\ref{eigenvalue}) by an analytic continuation procedure in the
mass eigenvalue $m = i \mu$, justified by the absence of any poles
in the complex $m$ plane except for the Kaluza-Klein poles.
Using the definition of the theta functions given in the Appendix,
we can rewrite the eigenvalue equation as :
\begin{eqnarray}
\frac{1}{\bar{m}^2}
& = & -  \int_{ {1/\Lambda^2}}^{\infty}  d t \
 e^{- \mu^2 t} \ \theta_3 \left(\frac{i t}{\pi R_1^2} \right) \theta_3 \left( \frac{i t}{\pi R_2^2} \right)
 \nonumber  \\
& = &
- \int_{R_c^2}^{\infty}
d t \ e^{- \mu^2 t} \
\theta_3 \left(\frac{i t}{\pi R_1^2} \right) \
\theta_3 \left( \frac{i t}{\pi R_2^2} \right)
\nonumber
\\
&&
- \pi R_1 R_2
\int_{ {1/\Lambda^2}}^{R_c^2}
\frac{d t}{t}
\ e^{- \mu^2 t} \
\theta_3 \left(\frac{i \pi R_1^2}{t} \right)
\ \theta_3 \left( \frac{i \pi R_2^2}{t} \right) \ ,
     \label{integral}
\end{eqnarray}
where we have introduced an UV cut-off $\Lambda$ noting that the integral
is log divergent. In the following equations we consider for
simplicity equal radii $R_1=R_2 = R_c$,  where $R_c$ defines
the compactification scale.
We also rewrote, using properties of the $\theta_3$ function defined in the
appendix, the last contribution in a more convenient way.

\textit{Large Radii Limit:} In the large radii limit, with respect to the
cutoff $\Lambda$ and by anticipating that the lightest mass eigenvalue
$m < R^{-1}$ (we will discuss in section~3 the solutions corresponding to the massive KK excitations), the approximate solution of the above eq.(\ref{integral}) becomes
\begin{equation}
    \label{integral2}
{1 \over \bar{m}^2} = {4 \pi^2 R_c^2 \over h_2} \simeq -  \pi
R_c^2 \ln({\Lambda^2 R_c^2}) + {1 \over m^2} \ .
\end{equation}
The second piece of the RHS of the above equation is the one corresponding to
$k_i = 0$: it is the infrared contribution coming from the integration region
$ R_c^2 < t < \infty$, whereas the first piece of the RHS, logarithmically divergent,
is coming from the ultraviolet region of heavy states
$ 1/\Lambda^2 < t < R_c^2 $. The first interpretation of the
result is as follows.  When the radii are small,
we have the ordinary ``volume suppressed mass" given by :
\begin{equation}
m^2_{\rm naive} = {h_2 \over 4 \pi^2 R_c^2 } \ .
\end{equation}
On the other hand
for large radii the logarithmic contribution from the UV has a significant
effect and we get the renormalised contribution
\begin{equation}
{1 \over m^2} = 4 \pi^2 R_c^2 \left( {1 \over h_2} + {1 \over 4
\pi} \ln({\Lambda^2 R_c^2}) \right) \ .
\end{equation}
This corresponds to the lightest eigenvalue. There is also a tower of massive KK states, the corresponding masses and their running will be presented in Section~3~(see~eq~(\ref{rd8})).  

Let us now move to understand the corresponding wave function. This wave function is easily obtained from the components of $\Psi_m$ in the $| k_1,k_2\rangle_{(k_1,k_2)\in \mathcal{I}}$ basis:
\begin{equation}
\varphi_{m} ({\bf y}) = \langle y_1, y_2 | \Psi_m \rangle =
\sum_{(k_1,k_2) \in \mathcal{I}} \langle y_1,y_2|k_1,k_2 \rangle
\langle k_1,k_2|\Psi_m \rangle \ .
\end{equation}
By using the eq.(\ref{egvec}),
this is evaluated to be:
\begin{equation}
    \label{wvfn}
\varphi_{m}({\bf y}) =
\frac{\mathcal{N}}{\sqrt{8 \pi^2 R_1 R_2}}
\sum_{k_1,k_2 = -\infty}^\infty
\frac{\cos(k_1 y_1/R_1 + k_2 y_2/R_2)}{m^2 - k_1^2/R_1^2 - k_2^2/R_2^2 }
\ ,
\end{equation}
where, again, we used the even properties of the cosine functions to transform the sums into sums from
$k_i=-\infty$ to $+\infty$.

The same results can actually be obtained in a simpler way through
appropriate boundary conditions, with special care in the fixed point where the localized
mass lives.

%%%%%%%%%%%%%%%%%%%%%%%%%%%%%%%%%%%%%%%%%%%%%%%%%%%%%%%%%%%%%%5
\subsection{Understanding through Boundary Conditions}
%%%%%%%%%%%%%%%%%%%%%%%%%%%%%%%%%%%%%%%%%%%%%%%%%%%%%%%%%%%%

As in 5D where we can reformulate orbifold projections in the presence of
boundary operators in terms of boundary conditions for fields
propagating in the fundamental domain~\cite{BCs}, we would like here to show that
we can derive the eigenvalue equation by imposing suitable boundary conditions in 6D.

In absence of any operator localized on the fixed points, the
$\mathbb{Z}_2$ parity assignment for
an even scalar field translates into the BCs at the fixed points
\begin{equation}
\partial_{1} \Phi \ = \ \partial_{2} \Phi \ = \  0 \ . \
\end{equation}
In the presence of the localized coupling $h_2$, the corresponding BCs can be read from the full equation of motion~(\ref{sc1}), which, upon the KK decomposition
$\Phi (x,{\bf y}) = \varphi_{m} ({\bf y}) \ \phi_m (x)$, $m$ being the 4D KK mass
$\partial_\mu \partial^\mu \phi_m (x) + m^2  \phi_m (x) = 0$, reads
\be
     \label{bc1}
( \partial_1^2 + \partial_2^2  \ + \ m^2 ) \ \varphi_{m} ({\bf y})
\ = \ h_2 \ \varphi_{m} ({\bf y})\ \delta^2 ({\bf y})
 \ .
\ee
To obtain the corresponding BCs at the origin, we
first regularize the mass distribution by distributing it along a
circle of radius $\epsilon$ around the origin.
The mass density is related to the localized dimensionless parameter
$h_2$ by
\begin{equation}
h_2 = 2 \pi \, \epsilon\,   m_\epsilon^2 \ .
\end{equation}
With this linear mass distribution, the BCs along the surface
$y_1^2+y_2^2=\epsilon^2$ ($n$ is the normal to the surface) becomes
\begin{equation}
\partial_n \varphi  =  \frac{1}{2} \, m_\epsilon^2 \, \varphi \ \label{regu} 
\end{equation}
and when $\epsilon$ shrinks to zero, we obtain the new BCs at the origin
\begin{eqnarray}
    \label{eq:newBCs}
\partial_{y_1} \varphi  = \frac{1}{2} \, h_2  \ \varphi \,\delta (y_2)
\ , \nonumber \\
\partial_{y_2} \varphi = \frac{1}{2} \, h_2  \ \varphi \, \delta (y_1) \ .
\end{eqnarray}

We will now check that these BCs lead to the eigenvalue equation~(\ref{eigenvalue}) derived earlier by the
 diagonalization of an infinite mass matrix. First, we note that the most general solution of the equation
 of motion~(\ref{bc1}) in the bulk takes the form
\be
    \label{eq:bulksol}
\varphi_{m}({\bf y}) = \mathcal{N'}
\sum_{k_1,k_2 = -\infty}^\infty
\frac{\cos(k_1 y_1/R_1 + k_2 y_2/R_2 + \alpha_{\bf k} )}{m^2 - k_1^2/R_1^2 - k_2^2/R_2^2}
\ee
which verifies
\begin{equation}
(\partial_1^2 + \partial_2^2) \varphi_{m} = -m^2  \varphi_{m}
+ \mathcal{N'} \sum_{k_1,k_2 = -\infty}^\infty \cos(k_1 y_1/R_1 + k_2 y_2/R_2 +\alpha_{\bf k} )
\ .
\end{equation}
In (\ref{eq:bulksol}), $\alpha_{\bf k}$ are phases related to the position of the localised
mass term. For a mass term localised at the origin, boundary conditions fix $\alpha_{\bf k}
= 0 $.  
Then, by using the identity $\sum_{k=-\infty}^{\infty} \cos (k y /R) =
2 \pi R \ \delta(y)$, we finally obtain
\begin{equation}
    \label{eq:bc2}
(\partial_1^2 + \partial_2^2) \varphi_{m} = -m^2  \varphi_{m}
+ 4 \pi^2 R_1 R_2  \ \mathcal{N'} \ \delta^2 ({\bf y}) \
\ . 
\end{equation}
The comparison of the boundary terms in~(\ref{bc1}) and (\ref{eq:bc2}) leads to the desired 
eigenvalue equation~(\ref{eigenvalue}). If the mass distribution is regularised according
to (\ref{regu}), then the sum over the KK integers $(k_1,k_2)$ in
(\ref{eigenvalue}) is effectively cut for large
momenta $|k_i| < k_{i,max} = R_i / \epsilon$. The size $\epsilon$ is related to the UV cutoff
defined in the previous section by $\Lambda = 1 / \epsilon$.

Alternatively, one can also derive the same eigenvalue equation by simply enforcing the modified BCs~(\ref{eq:newBCs}). To this end, one needs to massage the expression of the wavefunction~(\ref{wvfn}). Actually, one of the sum can be performed  by using the formula:
\be
\sum_{k =
-\infty}^{\infty} { \cos(kx) \over k^2 + \alpha^2} = {\pi \over
\alpha} {\cosh \alpha(\pi - x) \over \sinh \alpha \pi} \ .
\ee
We obtain:
 \bea
     \label{sumk1wvfn}
\varphi(y_1,y_2) &=& - \mathcal{N'} \pi R_1 R_2 
\sum_{k_2=-\infty}^{\infty}  
\frac{e^{ik_2 y_2/R_2}}{\sqrt{k_2^2 - m^2 R_2^2 }} 
\frac{\cosh \left( (R_1/R_2) \sqrt{ k_2^2 - m^2 R_2^2} \ (\pi - y_1/R_1) \right)}{ \sinh 
\left( (\pi R_1/R_2) \sqrt{k_2^2 - m^2 R_2^2} \right)} \ ,
\eea
which leads to the expression:
\begin{equation}
\partial_{y_1} \varphi ( y_1=0, y_2) = 2 \pi^2 R_1 R_2 \ \mathcal{N'} \
\delta (y_2) \ .
\end{equation}
Inserting this relation into the BCs at the origin, we are left again with the eigenvalue
equation~(\ref{eigenvalue}). The boundary condition in the second direction is 
satisfied in the same way.

%%%%%%%%%%%%%%%%%%%%%%%%%%%%%%%%%%%%%%%%%%%%%%%%%%%%%%%%%%%%%%%%%%%%%%%%%55
\section{Regularization Dependence and running}

The interpretation of the logarithmic divergence in the sum is actually
rather clean, by realizing that the brane localized couplings,
following~\cite{ggh}, do run in the sense of the four-dimensional
renormalisation group. Our example
is similar in spirit to the one analysed in~\cite{gw}.
It was shown there that the (euclidean) Dyson resummation of the scalar
propagator, by including the brane localized mass insertions, in a mixed representation :
4d spacetime momentum space and 2d extra-dimensional coordinate space, is
\begin{eqnarray}
 G (q_4,{\bf x} , {\bf y} ) 
 & = &   D (q_4,{\bf x} , {\bf y}) - h_2 \ D
(q_4,{\bf x} ,{\bf 0} )  D (q_4,{\bf 0} , {\bf y}) 
 \nonumber
\\
&&
\ \ +
h_2^2 \   D (q_4,{\bf x} , {\bf 0}) \ D (q_4,{\bf 0},{\bf 0})   D
(q_4,{\bf 0} , {\bf y}) +  \cdots \ \nonumber
\\
& = &  D (q_4,{\bf x} , {\bf y})  - {h_2 \over 1 + h_2  
D (q_4,{\bf 0},{\bf 0})} \ 
 D (q_4,{\bf x} , {\bf 0})  D (q_4,{\bf 0} , {\bf y}) \ , \
    \label{rd1}
\end{eqnarray}
where $D (q_4,{\bf x} , {\bf y})$ is
the free scalar six-dimensional propagator of four-momentum $q_4$
and two-dimensional positions  ${\bf x} , {\bf y}$. The propagator at the origin at the compact 
space sum in the compact space $D (q_4,{\bf 0},{\bf 0})$
is defined via
\begin{equation}
 D (q_4,{\bf 0},{\bf 0}) \ = \ \sum_{\bf k} D (q_4,{\bf k}) \ \equiv \ {1 \over 4
\pi^2 R_1 R_2} \sum_{k_1,k_2} {1 \over q_4^2 + k_1^2/R_1^2 +
k_2^2/R_2^2} \ , \
    \label{rd2}
\end{equation}
where  ${\bf k}= (k_1,k_2)$ are the Kaluza-Klein
momenta in the transverse space.
The pole of the resummed
propagator~(\ref{rd1}) defines the physical four-dimensional
masses $q_4^2 = m^2$. One therefore gets, by going back in
the Minkowski space
\begin{equation}
    \label{propagator}
{4 \pi^2 R_1 R_2 \over h_2} =
\sum_{k_1,k_2 = - \infty}^{\infty} {1 \over m^2 -
  k_1^2/R_1^2 -  k_2^2/R_2^2} \ ,
    \label{rd3}
\end{equation}
which is nothing that the eigenvalue equation we derived
earlier. The off-shell expression~(\ref{rd1}), however, is more
suitable for renormalisation purposes since by standard techniques
it leads, for $\mu > 1/R_c$ (in the case of equal radii, $R_1=R_2=R_c$, that we are considering here), 
to the renormalisation group flow of
the $h_2$ coupling\footnote{Ref.~\cite{gw} finds a $\beta$ function $1/(4\pi)$ for the orbifold 
($\alpha=1/2$). The factor 2 of discrepancy with our result is coming from the different normalisation 
of the brane-localized coupling. We did normalise it on the torus, whereas Ref.~\cite{gw} normalises 
it on the orbifold. The rescaling $h_2 \to h_2/2$ exactly accounts for the missing factor of 2 in the 
$\beta$ function.}
 \begin{equation}
  \mu { d h_2 \over d \mu } = {h_2^2 \over 2
\pi} \ .
    \label{rd4}
\end{equation}
Indeed, an approximate evaluation of the
propagator for different values of the four-dimensional momentum
gives
\begin{eqnarray}
&& \sum_{\bf k} D (q_4,{\bf k})  \sim {1 \over 4 \pi}
\ln {\Lambda^2 \over q_4^2} \quad , \quad {\rm for} \ \ q_4^2 >
1/R_c^2 \ , \nonumber
\\
&& \sum_{\bf k} D (q_4,{\bf k})  \sim {1 \over 4 \pi} \ln
(\Lambda^2 R_c^2) + {1 \over 4 \pi^2 R_c^2 q_4^2} \quad , \quad 
{\rm for} \ \ q_4^2 < 1/R_c^2 \ , 
	\label{rd04}
\end{eqnarray}
and therefore for
$q_4^2 > 1/R_c^2$ the propagator exhibits a logarithmic scale dependence
typical of a renormalisation group running of the coupling $h_2$.
The coupling $h_2$ in the eigenvalue equation~(\ref{rd3}) is
interpreted as the value of the coupling at a high cutoff scale
$\Lambda$, naturally of the order of the mass scale
(higher-dimensional Planck mass or the string scale) of the
microscopic theory. Following this interpretation, the
eigenvalue equation~(\ref{rd3}), solved under the physically
relevant, for our case, situation $m \ll 1/R_i$ (we assume
$R_1 \sim R_2$ for simplicity), defines the pole mass. For energy
scales $\mu > 1/R_{1,2}$, the $q_4^2$ dependence can be renormalised away by an appropriate 
counter-term, which gives rise to the running of the mass eigenvalue:
\begin{equation}
    \label{rge}
{1 \over m^2 (\mu)} = {4
\pi^2 R_1 R_2 } \left( {1 \over h_2} + {1 \over 2
  \pi} \ln {\Lambda \over \mu} \right) \equiv {4 \pi^2 R_1 R_2 \over
h_2 (\mu)} \ .
    \label{rd5}
\end{equation}

The eigenvalue equation~(\ref{rd3}) has other solutions
corresponding to the massive Kaluza--Klein states. They can be easily
computed in the leading approximation in $h_2 \ll 1 $ by searching solutions
of the form
\begin{equation}
m^2_{(l_1,l_2)} = 
\left( \frac{l_1^2}{R_1^2} + \frac{l_2^2}{R_2^2} \right) 
(1 + \epsilon_{l_1,l_2} ) \quad , \quad \epsilon_{l_1,l_2} \ll1 \ . 
	\label{rd6}
\end{equation}
A straightforward computation gives
\begin{equation}
m^2_{(l_1,l_2)} \simeq 
\frac{l_1^2}{R_1^2} + \frac{l_2^2}{R_2^2} 
+ \frac{c_{l_1,l_2}}{ 4 \pi^2 R_1 R_2} 
\ \left( \frac{1}{h_2} + \frac{1}{4\pi} \ln \frac{\Lambda^2}{l_1^2/R_1^2+l_2/R_2^2}
\right)^{-1}
 \ .
    \label{rd7}
\end{equation}
$c_{l_1,l_2}$ is a degeneracy factor counting the number of solutions to the integer equation
$l_1^2/R_1^2+l_2^2/R_2^2=k_1^2/R_1^2+k_2^2/R_2^2$, with $k_1,k_2=-\infty\ldots +\infty$. For instance, in the case of equal radii,
$c_{1,1}=4$ but $c_{0,10}=12$.

Similarly to the lightest eigenvalue, the KK masses also run, but
in this case the running occurs above their pole mass $\mu^2 >
m^2_{(l_1,l_2)}$
\begin{equation}
m_{(l_1,l_2)}^2 (\mu) \simeq  \frac{l_1^2}{R_1^2} + \frac{l_2^2}{R_2^2} 
\ + \ c_{l_1,l_2} \, 
\frac{h_2 (\mu)}{4 \pi^2 R_1 R_2} \ .
    \label{rd8}
\end{equation}

%%%%%%%%%%%%%%%%%%%%%%%%%%%%%%%%%%%%%%%%%%%%%%%%%%
\section{The Fermionic Case}
%%%%%%%%%%%%%%%%%%%%%%%%%%%%%%%%%%%%%%%%%%%%%%%%%%

Before deriving the mass formula, let us first describe our
framework in more detail.  We will assume that the Standard Model
fermions are restricted to the~4D brane whereas a singlet~6D
Weyl neutrino is free to propagate in the full six dimensional
space\footnote{In the presence of gravity or bulk gauge symmetries, anomaly cancellation
in six dimensions gives nontrivial constraints on the 6D Weyl fermionic spectrum, which have to be taken
into account in building a complete theory. }. The generalization to more than one singlet neutrino,
needed in the three generation case, is straightforward. The six
dimensional bulk action is simply
\begin{equation}
    \label{6dlag}
S \ = \
\int d^4x d^2y \ i \bar{\Psi} \Gamma^M \mathcal{D}_M \Psi
\ .
%+ \int d^4x \ \ \left[ h {\bar \nu}_L \Psi (y_1 = y_2 =0) \ H + h.c. \right] \ ,
\end{equation}
Notice that no Dirac nor Majorana bulk mass is allowed for a 6D Weyl fermion.  

The $\mathbb{Z}_2$ orbifold projection acts on $\Psi$ as
\begin{equation}
\mathbb{Z}_2 \ \Psi ({\bf y}) \ = \ i \ e^{-(\pi/2) \, \Gamma^5 \Gamma^6} \, \Psi (- {\bf y}) \ .
\end{equation}
Decomposing the eight-component 6D Weyl spinors into two four-component 4D Weyl spinors (see Appendix~B), the $\mathbb{Z}_2$ action becomes
\begin{equation}
\Psi =
\left( \begin{array}{c} \lambda_1 \\  \lambda_2 \end{array} \right), \ \ \ \
\mathbb{Z}_2 \lambda_1 ({\bf y}) = \lambda_1 (-{\bf y}), \,
\mathbb{Z}_2 \lambda_2 ({\bf y}) = - \lambda_2 (-{\bf y}) \ .
\end{equation}
This allows us to write a brane localized coupling between the SM left-handed neutrino and the
even fermion $\lambda_1$:
\begin{equation}
 \int d^4x \ \ \left( h\,  {\bar \nu}_L \lambda_1 ({\bf y} =0) \, H + h.c. \right) \ ,
\end{equation}
where $H$ is the 4D Higgs field and $h$ is the Yukawa coupling. The mass dimensions of $\Psi, \nu_L, H$ are respectively $5/2, 3/2$ and $1$.

Written in two-component spinor notations, the 6D lagrangian  takes the form :
\begin{eqnarray}
    \label{6dlag2comp}
\mathcal{L} &=&
- i \lambda_1~ \sigma^\mu~ \partial_\mu \bar{\lambda}_1
-
i \lambda_2~ \sigma^\mu~ \partial_\mu \bar{\lambda}_2
+
\lambda_1 \left( \partial_5 + i \partial_6 \right) \lambda_2
 \nonumber \\
& &
- \bar{\lambda}_2 \left(  \partial_5 - i \partial_6 \right) \bar{\lambda}_1
+  g_2 (\nu_L \lambda_1 + \bar{\nu}_L \bar{\lambda}_1) \delta ^2 ({\bf y}) \ ,
\end{eqnarray}
where we have now introduced
\begin{equation}
g_2 = h \langle H \rangle
\end{equation}
the brane-localized `Dirac' neutrino
mass parameter which actually, in analogy with the scalar case, is a
dimensionless parameter. We will study the
phenomenology for two cases: {\it (a)} Dirac type neutrinos;
{\it (b)} Majorana type neutrinos. Before proceeding further, we would like to
point out that the equations of the motion are given by :
\begin{eqnarray}
&
i \sigma^\mu \partial_\mu \bar{\lambda}_1 - (\partial_5 + i \partial_6) \ \lambda_2
=  - g_2 \ \nu_L \delta^2 ({\bf y}) \ ,
\nonumber
\\
&
i \sigma^\mu \partial_\mu \bar{\lambda_2} +
(\partial_5 + i \partial_6) \ \lambda_1
=  0 \ ,
\nonumber
\\
&
i \sigma^\mu \partial_\mu \bar{\nu_L}
 =  - g_2 \ \lambda_1 ({\bf y}=0) \ .
\end{eqnarray}
Using the identity on the Pauli matrices given in the Appendix~B, one can easily combine these coupled  first order differential equations to obtain an uncoupled second order differential equation that is nothing but the  Klein--Gordon equation  in 6D:
\begin{equation}
\left( \partial_\mu \partial^\mu - \partial_5^2 -\partial_6^2 \right) \,  \lambda_1 (x,{\bf y})
+
g_2^2  \ \lambda_1 (x,{\bf y}) \ \delta^2({\bf y})=0 \
\end{equation}
and therefore our fermionic problem with brane localized Dirac
mass term is reduced to the one of the bulk scalar field with
brane localized mass term studied in the previous Sections.

%%%%%%
\subsection{Dirac Neutrinos}
%%%%%%

Let us now study the mass matrix for the case of Dirac neutrinos.
To this end we will follow the procedure first developed in the scalar
case and we will expand the 6D fermion $\Psi$ on a complete set of
functions. We first write the 6D Weyl fermion in terms of a pair 
of two-component spinors, $\lambda_{1,2}$. According to their $\mathbb{Z}_2$ parities,  $\lambda_{1,2}$
can be decomposed as:
\begin{eqnarray}
& \displaystyle
\lambda_1 (x,{\bf y}) =
\frac{1}{\sqrt{2 \pi^2 R_1 R_2 }}
\sum_{(k_1,k_2) \in \mathcal{I}}
\frac{\cos (k_1 y_1/R_1 + k_2 y_2/R_2)}{\sqrt{2^{\delta_{k_1 0}\, \delta_{k_2 0}}}}
\ \chi_{(k_1,k_2)} (x) \ ,
\\
& \displaystyle
\lambda_2 (x,{\bf y}) =
\frac{1}{\sqrt{2 \pi^2 R_1 R_2 }}
\sum_{(k_1,k_2) \in \mathcal{I}}
\sin (k_1 y_1/R_1 + k_2 y_2/R_2)
\ \xi_{(k_1,k_2)} (x) \ .
\end{eqnarray}
After integration over the two extra dimensions, the mass terms in the action become
\begin{equation}
    \label{6dlagdirac}
\mathcal{L}_{\rm mass}
=
\sum_{(k_1,k_2) \in \mathcal{I}}
\left( \frac{k_1}{R_1} + i \frac{k_2}{R_2} \right)
\chi_{(k_1,k_2)} \xi_{(k_1,k_2)}
+
\bar{m}
\sum_{(k_1,k_2) \in \mathcal{I}}
\frac{\sqrt{2}}{\sqrt{2^{\delta_{k_1 0}\, \delta_{k_2 0}}}}
\nu_L \chi_{(k_1,k_2)}
\ + h.c.
\end{equation}
with
\begin{equation}
\bar{m} = \frac{g_2}{\sqrt{4 \pi^2 R_1 R_2}}\ .
\end{equation}
We first notice that the phases in the KK masses
can be absorbed by redefining the fields $\xi_{(k_1,k_2)}$, while the phase in $\bar{m}$ can be removed by a redefinition of $\nu_L$.
Thus, these phases do not have any
physical effects and all the mass terms can be taken real. Next it is convenient to introduce the combinations (by convention, we define $A_{(0,0)}= \chi_{(0,0)}$ and $B_{(0,0)}=0$)
\begin{equation}
A_{(k_1,k_2)} =
\frac{\chi_{(k_1,k_2)} + \xi_{(k_1,k_2)}}{\sqrt{2}}
\ \ {\rm{and}}
\ \
B_{(k_1,k_2)} =
\frac{\chi_{(k_1,k_2)} - \xi_{(k_1,k_2)}}{\sqrt{2}}  \ .
\end{equation}
The mass terms now write
\begin{equation}
\frac{1}{2}
\left( \nu_L \, A_{(k_1,k_2)} B_{(k_1,k_2)} \right)
\mathcal{M}
\left(
\begin{array}{c}
\nu_L\\
A_{(p_1, p_2)}\\
B_{(p_1, p_2)}
\end{array}
\right)
\end{equation}
with
\begin{equation}
\mathcal{M}=
\left(
\begin{array}{ccc}
0 & \bar{m}
& \bar{m}
\\
\bar{m}
&\tv{20} \sqrt{\frac{k_1^2}{R_1^2}+\frac{k_2^2}{R_2^2}} \delta_{k_1 p_1} \delta_{k_2 p_2} & 0
\\
\bar{m} &0
& - \sqrt{\frac{k_1^2}{R_1^2}+\frac{k_2^2}{R_2^2}} \delta_{k_1 p_1} \delta_{k_2 p_2}
\end{array}
\right) \ .
\end{equation}
The mass matrix $\mathcal{M}$ is real and symmetric.
We will diagonalize is in a similar
manner as we did for the scalar mass squared matrix.
The characteristic equation defining the eigenvector $\Xi_m$ associated with the eigenvalue
$m$ is
\be
    \label{6dcharacter}
\mathcal{M} \ \Xi_m \ = \ m \ \Xi_m \ .
\ee
This matrix equation is equivalent to the infinite set of equations
\begin{eqnarray}
    \label{egvec1}
& \displaystyle
\bar{m} \sum_{\mathcal{I}}
\left( \langle A_{(k_1,k_2)} | \Xi_m \rangle + \langle B_{(k_1,k_2)} | \Xi_m \rangle \right)
= m  \langle \nu_L | \Xi_m \rangle \ ,
\\
    \label{egvec2}
& \displaystyle
\bar{m} \langle \nu_L | \Xi_m \rangle
+ \sqrt{\frac{k_1^2}{R_1^2} + \frac{k_2^2}{R_2^2} } \langle A_{(k_1,k_2)} | \Xi_m \rangle
= m \langle A_{(k_1,k_2)} | \Xi_m \rangle \ ,
\\
    \label{egvec3}
& \displaystyle
\bar{m} \langle \nu_L | \Xi_m \rangle
- \sqrt{\frac{k_1^2}{R_1^2} + \frac{k_2^2}{R_2^2} } \langle B_{(k_1,k_2)} | \Xi_m \rangle
= m \langle B_{(k_1,k_2)} | \Xi_m \rangle \ .
\eea
Plugging back in eq.~(\ref{egvec1}) the expressions for $\langle A_{(k_1,k_2)} | \Xi_m \rangle$ and
$\langle B_{(k_1,k_2)} | \Xi_m \rangle$ obtained from eqs.~(\ref{egvec2}--\ref{egvec3}), we finally obtain the desired eigenvalue equation
\begin{equation}
\frac{1}{\bar{m}^2}\  
= 
\sum_{k_1,k_2 = -\infty}^{\infty} 
\frac{1}{m^2 - k_1^2/R_1^2 - k_2^2/R_2^2 } \ .
\end{equation}
%

%%%%%%
\subsection{Majorana Case}
%%%%%%

We now discuss the case of Majorana neutrinos. To the lagrangian
discussed above~(\ref{6dlag2comp}) we want to add lepton number
violating Majorana mass terms involving the Kaluza--Klein states
of the bulk fermion~$\Psi$. Note that there are two ways to
break the lepton number : {\it (a)} Breaking the lepton number on the brane
by introducing on the brane a lepton number violating mass term for the singlet neutrino;
{\it (b)} Breaking the lepton number in the bulk
through a bulk Majorana mass for the singlet neutrino. We will
directly present the mass spectrum of these two cases in the
following. Let us first study the case {\it (a)} and let us add the
lepton number violating mass term $M_0$ on the brane
\begin{equation}
\int d^4 x \ d^2 y \
( M_0 \ \lambda_1 \lambda_1 + h.c )  \ \delta^2 ({\bf y}) \ .
\end{equation}
Notice that actually from a 6D perspective $M_0$ has mass
dimension $-1$, whereas after the KK expansion the physical mass
parameter is $M_0 / (4 \pi^2 R_1 R_2) $. Similar to the Dirac mass
considered previously, all phases in the Kaluza--Klein complex
masses can be redefined away and have no physical meaning. By a
straightforward generalisation of the previous diagonalisation, we
find the eigenvalue equation
\begin{equation}
\frac{1}{\bar{m}^2 + m M_0/(4 \pi^2 R_1 R_2)}\  =
\sum_{k_1,k_2 = -\infty}^{\infty}
\frac{1}{m^2 -k_1^2/R_1^2 - k_2^2/R_2^2 }
\ .
\end{equation}
By considering again for
simplicity the case of two equal radii $R_1=R_2=R_c$ and
evaluating as before the double sum by keeping the leading IR and
UV contributions, we find in the large radii limit
\begin{equation}
\frac{1}{m^2} = 4 \pi^2 R_c^2
\left( \frac{1}{g_2^2 + m M_0} + \frac{1}{4 \pi} \ln (\Lambda^2 R_c^2) \right) \ .
\end{equation}
The natural interpretation is again in terms of the running of the physical mass
\begin{equation}
\frac{1}{m^2 (\mu)} \ = \
\frac{ 4 \pi^2 R_c^2}{( g_2^2 + m M_0) (\mu)} \ .
\end{equation}
The case of the
brane Majorana mass is the simplest but also the most problematic,
since due to the double volume suppression the lepton number
violation, is small. The case of the bulk Majorana mass $M$ is more
subtle. First of all, a standard Majorana, Lorentz invariant mass
in 6D, ${\bar \Psi}^C \Psi$ cannot be written for a Weyl fermion,
since it mixes 6D Weyl fermions of opposite
chiralities\footnote{E.D. thanks P. Holstein and S. Lavignac for
discussions on this issue.}. On the other hand, a Majorana mass term involving only a
6D Weyl fermion, of the form ${\bar \Psi}^C \Gamma_6 \Psi$ can be
written, at the expense of breaking the 6D Lorentz symmetry
which could be considered as being spontaneously generated by the
vev of some vector field. In this case, it is not possible anymore
to eliminate the phases in the KK masses by field redefinitions,
even for real Majorana mass. Interestingly enough, the interplay
between Kaluza--Klein masses and bulk Lorentz violating Majorana
mass generates CP violation. This observation could be related to
previous proposals to relate the CP symmetry to discrete subgroups of a
higher-dimensional Lorentz group~\cite{ckn}. The eigenvalue equation in this
case is
\begin{equation}
\frac{g_2^2}{ 4 \pi^2 R_1 R_2} \ 
\sum_{k_1,k_2 = -\infty}^{\infty}
\frac{1}{ m - M - i k_1/R_1 + k_2/R_2 }
\ = \ m \ .
\end{equation}

%%%%%%%%%%%%%%%%%%%%%%%%%%%%%%%%%%%%%%%%%%%%%%%%%%%
\section{Neutrino oscillations}
%%%%%%%%%%%%%%%%%%%%%%%%%%%%%%%%%%%%%%%%%%%%%%%%%%%

Generally, the gauge neutrino eigenstates $\nu_f$ are related to
the set of mass eigenstates ${\tilde \nu}_i$ through a unitary
mixing matrix $U$ as \be \nu_f = \sum_i U_{fi} {\tilde \nu}_i \ ,
\label{no1} \ee where the matrix $U$ is extracted from the
neutrino mass matrix. The probability of oscillation between two
gauge eigenstates $\nu_f$ and $\nu_f'$ after a time $t$ is given
by 
\begin{equation}
P_{f \rightarrow f'} (t) = 
\sum_i \left| U_{fi} U_{f'i} \right|^2 
+ 2 \sum_{i > j} {\rm Re} \left( U_{fi} U^*_{f'i} U^*_{fj} U_{f'j} e^{[i (E_j-E_i)t]} \right) \ , 
    \label{no2}
\end{equation}
where $E_i = \sqrt{p^2 + m_i^2}$ is the energy of the mass
eigenstate ${\tilde \nu}_i$. Of a particular interest for our case
where the active neutrinos oscillates into bulk states is the
survival probability after a time $t$, given by 
\begin{equation} 
P_{f\rightarrow f} (t) = 
\left| 
\ \sum_i 
|U_{fi}|^2 e^{i E_i t} \ \right|^2 
\ .
\ee 
We consider in the following the case of the Dirac neutrinos
discussed in detail in Section~4. The matrix $U$ in this case is
(doubly) infinite dimensional and, from~(\ref{egvec1})-(\ref{egvec3}) its explicit form is
\begin{eqnarray}
& \displaystyle 
\langle \Xi_{m_i} | U | \nu_L \rangle = \frac{1}{\sqrt{N_i}}\ ,
\\
&\displaystyle 
\langle \Xi_{m_i} | U | A_{(k_1,k_2)} \rangle = \frac{1}{\sqrt{N_i}}\, 
\frac{\bar{m}}{m_i-\sqrt{k_1^2/R_1^2+k_2^2/R_2^2}}\ ,
\ (k_1,k_2) \in \mathcal{I} \ ,
\\
&\displaystyle 
\langle \Xi_m | U | B_{(k_1,k_2)} \rangle = \frac{1}{\sqrt{N_i}}\,
\frac{\bar{m}}{m_i+\sqrt{k_1^2/R_1^2+k_2^2/R_2^2}}\ ,
\ (k_1,k_2) \in \mathcal{I}\backslash (0,0) \ .
	\label{no3} 
\end{eqnarray}
The unitarity property of the $U$ matrix determines the normalisation constant,  $N_i$, to be
\begin{eqnarray} 
N_i 
& = &
1+ \bar{m}^2 \sum_{k_1,k_2=-\infty}^\infty \frac{m_i^2 + k_1^2/R_1^2 + k_2^2/R_2^2}{
\left( m_i^2 - k_1^2/R_1^2 - k_2^2/R_2^2\right)^2}
\nonumber\\
& = &
 2 \bar{m}^2 m_i^2 \sum_{k_1,k_2=-\infty}^\infty \frac{1}{
\left( m_i^2 - k_1^2/R_1^2 - k_2^2/R_2^2\right)^2}
	\label{no4}
\end{eqnarray}
where the eigenvalue equation~(\ref{eigenvalue}) has been used to arrive at the last expression.
We notice that the last summation~(\ref{no4}) is
manifestly UV finite and therefore the only cutoff dependence arise implicitly through the running
of the physical neutrino masses.
The active neutrino
survival probability, in the non-relativistic approximation, is
then given by the formula~\cite{ddg2} 
\begin{equation}
P_{\nu_L \rightarrow \nu_L} (t) 
= 
\left| 
\sum_{(k_1,k_2)\in \mathcal{I}}  \frac {2}{N_{(k_1,k_2)}} \ 
e ^{i \frac{m_{(k_1,k_2)}^2 t}{2 p}} \ \right|^2 
=
\left| 
\sum_{k_1,k_2=-\infty}^{\infty}  
\frac {2^{\delta_{k_1 0} \delta_{k_2 0}}}{N_{(k_1,k_2)}} \ 
e ^{i \frac{m_{(k_1,k_2)}^2 t}{2 p}} \ \right|^2 
\ . 
	\label{no6} 
\end{equation}
The factor~2 in the first sum accounts for the fact that each massive state corresponds to 
a Dirac neutrino. An approximate estimation of the normalisation factors $N_{(k_1,k_2)}$ 
gives the result
\begin{equation}
%N_{(0,0)} \ \simeq \ {2 {\bar m}^2 \over m_0^2} \quad , \quad 
N_{(k_1,k_2)} 
%\ \simeq \   
%\frac{1}{2^{\delta_{k_1 0}} 2^{\delta_{k_2 0}}}
%\, \frac{8 \bar{m}^2 m_{(k_1,k_2)}^2}{\left(m_{(k_1,k_2)}^2 - k_1^2/R_1^2 - k_2^2/R_2^2 \right)^2}
\ \simeq \ 
c_{k_1,k_2} \
\frac{2 \bar{m}^2 m_{(k_1,k_2)}^2}{\left(m_{(k_1,k_2)}^2 - k_1^2/R_1^2-k_2^2/R_2^2\right)^2}
 	\label{no7}
\end{equation}
where 
$m_{(k_1,k_2)}$ is the pole mass eigenvalue given by the eq.~(\ref{rd7}) and
$c_{k_1,k_2}$ are the degeneracy factors defined in Section~3.
We can anticipate from~(\ref{no7}) some similarities and also some differences 
between the 5D and the present 6D case. Like in the 5D case~\cite{ddg2}, 
for large KK masses $N_i \sim m_i^2$ and the active neutrino mostly
oscillates into the lowest mass states. In the 6D case, on the other
hand, the degeneracy of the massive states is higher than in 5D and the 
decoupling of the massive states is slower than in 5D.  

%\begin{figure}[ht]
%\centerline{\includegraphics{plot3.eps}}
%\caption{The survival probability for an oscillating Dirac neutrino is shown 
%as a function of $t/(2 R^2 p)$. Here the (double) sum is over twenty states. 
%As the number of states increases, the $P(t)$ would reach 1. The coupling
%$g_2$ is here $\mathcal{O}(1)$, whereas the radii are chosen to be eV$^{-1}$.}
%	\label{fig:oscillations}
%\end{figure}

Numerically too, we find similar behaviour. For a relatively small number
of states, $P(t)$ is always smaller than $1$. In fact, even $P(0)$ is smaller
than 1, unless one sums over all the states. $P(t)$ oscillates with $t$,
though it is always remaining smaller than $P(0)$. The oscillations become more
rapid as we sum over more and more states. 

%In the Fig.~\ref{fig:oscillations}, we show a numerical
%example, where we have (double) summed over the first 20 states. The 
%probability is shown as a function of $t/(2 R^2 p)$. We have chosen natural
%values of $g_2$ to be $\mathcal{O}(1)$ and $R^{-1}$ to be of the eV scale. 

The brane Majorana case can simply be recovered from the previous
expressions by the replacement ${\bar m}^2 \rightarrow {\bar m}^2 + m_i M_0$.

%%%%%%%%%%%%%%%%%%%%%
\section{Phenomenology}
%%%%%%%%%%%%%%%%%%%%%

\begin{figure}[ht]
\centerline{\includegraphics[scale=0.4]{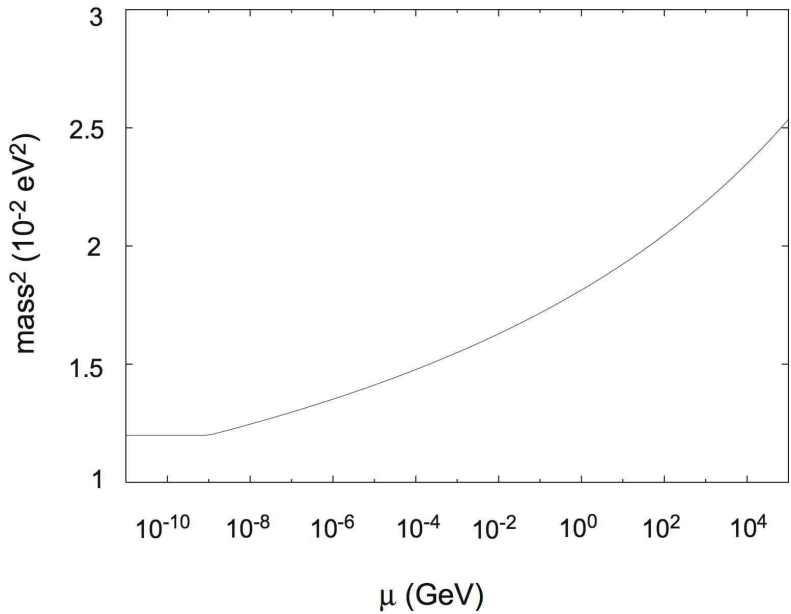}}
\caption{Running of the neutrino squared masses with the energy. We have chosen $\Lambda ~=~ 10^5$ GeV, $g_2~=~1$, and $R_1~=~R_2~ =~1$ eV$^{-1}$. The running occurs between the cutoff scale and the compactification scale. Below the compactification scale, there is not running anymore.}
		\label{fig:running}
\end{figure}

In this Section, we will detail some phenomenological consequences
which could be of interest for neutrino masses and mixings. First
of all, let us try to see the significance of the logarithmic
running with some realistic numbers. Fig.~\ref{fig:running} plots the running of the neutrino mass-squared with 
the energy. We have taken the radii to be of
the order of 1 eV$^{-1}$  with the fundamental Planck scale
$M_{\star}$ taken to be around 100~TeV, in order to recover the 4D Planck scale $M_P$ through the
standard relation $R_1 R_2 M_{\star}^4 = M_P^2 $~\cite{add1}.  Note that these
numbers are close to the present limits from cosmology on the size of two
large extra dimensions~\cite{sn87a}. The coupling $g_2$ is set to one at the
cutoff (microscopic) scale. We plot eq.~(\ref{rge}) and we see that
the neutrino mass  changes by
at least a factor of $3$ while running with the radius. Below
$R^{-1}$, there is no logarithmic running anymore and the masses keep
their values at the compactification scale. This explains the horizontal line
below $R^{-1}$. We interpret this result as follows. The pole of
the propagator as given in eq.~(\ref{propagator}), defines the
physical mass of the neutrino which is equivalent to the `running'
eigenvalue evaluated at $R^{-1}$. As long as neutrinos are
`on-shell', they carry this mass. The `running' neutrino masses on
the other hand varies with the off-shell momentum. Experimentally, the effect of
this `running' could be possibly seen in processes where neutrinos
are off-shell. The best probed process in this regard is
neutrinoless double beta decay which involves lepton number
violating neutrino masses.  We note only
that as long as the lepton number is violated by Majorana masses
and the running of the Dirac masses does occur, this is enough to
have observable effects due to the running between the pole mass
and the off-shell momentum, of the order the nucleon mass, in the
neutrinoless double beta decay diagrams. 
A detailed analysis of this case is beyond our goal here. 
On the other hand,
irrespective of whether lepton number is violated or not, this
running should always be considered while constructing models of
fermion masses in 6D. This is especially true in models where
large radii are considered. One can interpret that the `bare'
couplings in the lagrangian given at the scale $\Lambda$ would get
modified as they `run' towards their physical values. While most
of the present discussion has been concentrated for the case of a
single generation, we can easily extend the above analysis for
more than one flavour. First of all, it is useful to recall that in order to generate
three non-vanishing masses for the three active neutrinos, we need to introduce three
distinct bulk sterile Weyl fermions $\Psi_a$ and therefore the brane-bulk mixing mass term becomes a $3 \times 3$
matrix. The obvious change in the RGE evolution of the dimensionless coupling $h_2 \equiv g_2^2 $ is
\be
 \mu {d (h_2)_{ab} \over d \mu} \ = \ {1 \over 2 \pi}  \
 (h_2)_{ac}
 (h_2)_{cb}\ ,
\ee
whereas the $3 \times 3$ lightest eigenvalue mass matrix has the RGE evolution
\be
  \biggl({1 \over m^2 (\mu)}\biggr)_{ab} \ = \ {4 \pi^2 R_1 R_2 } \left[ \biggl({1 \over h_2}\biggr)_{ab} + {1 \over 2
  \pi} \ \delta_{ab}
\ln {\Lambda \over \mu} \right] \equiv {4 \pi^2 R_1 R_2 } \ 
\biggl( { 1 \over h_2 (\mu)} \biggr)_{ab} \ .
\ee
The classical running then modifies the flavour
mixing present at the scale $\Lambda$ by renormalising it in evolving towards low energy.
Since in order to have sizable running the active neutrinos are
required  not to be
much lighter than the sterile neutrinos, the oscillations into the
bulk sterile states are constrained by the existing experimental data. 
The two flavour active-sterile oscillations can be suppressed by choosing
a sufficiently small coupling $h_2$. For example, for $h_2~=~0.05$ we find
(for $\Lambda = 100$~GeV and summing up to first 200 states\footnote{For
these values, the running of the neutrino masses from the cut-off scale
to $R^{-1}\sim$~eV could be as large as 16 \%.}), that the
survival probability of the active neutrinos can stay around 70\%.
The survival probability obviously decreases as $h_2$ increases. 
 When
other active flavours are added, since mass differences for active 
neutrinos $\Delta m^2_{\nu_e \nu_{\mu}}, \Delta m^2_{\nu_e \nu_{\tau}} \ll 
 R_c^{-2} \sim eV^2$, for small values of the coupling $g_2$ and for all KK 
states the mass differences 
$m_{\nu_i}^2- m_{KK}^2 \gg \Delta m^2_{\nu_e \nu_{\mu}} , 
\Delta m^2_{\nu_e \nu_{\tau}}$. In this case, the oscillations into sterile 
neutrinos are subdominant with respect to active neutrino oscillations.
Obviously, the above results are qualitative and a full phenomenological 
analysis is required. We postpone it for a later work. 

  It is more useful to write the running neutrino mass directly as a
  function of the physical pole mass $m_{\nu}^2$. In the case of Dirac
  neutrinos and for one generation, for simplicity,
  we get
\be 
m_{\nu}^2 (\mu) \ = \ m_{\nu}^2 \ \frac{1 + {g_2^2 \over 2 \pi} \ln
  (\Lambda R_c)}{1 + {g_2^2 \over 2 \pi} \ln ({\Lambda \over \mu})} \
. \label{ph1} 
\ee 
This running is a small effect for small
couplings $g_2$. For example, for radii $R_c^{-1} \sim {\rm eV}$ and in
the normal hierarchical neutrinos scenario, the coupling $g_2$ for the electron
neutrino has to be of order $10^{-2}$. In this case, the running
(\ref{ph1}) in the energy range eV and GeV is too small to
have observable effects. In the inverse hierarchy (or almost
degenerate) neutrinos scenario, however, for $g_2 \sim \mathcal{O}(1)$, the
effect is large, as already stressed. This fits well with the fact
that neutrinoless double beta decay amplitude, proportional to the effective Majorana
mass
\be
m_{ee} \ = \ \sum_a U_{ea}^2 m_a \ , 
\ee
where $a$ are the three active neutrino states,  
increases with the absolute value of the neutrino
masses.

The question then remains about other physical effects of the sterile neutrinos. For the
large radii, $R_c \sim \mathcal{O}({\rm eV}^{-1}) - \mathcal{O}({\rm keV}^{-1})$, 
we should expect a large number of very light sterile states. The effects of the
virtual sterile states in loop processes involving neutrinos, in
astrophysics and cosmology need a separate careful analysis, which is very important 
for the validity of the models we proposed in this paper. This is beyond the goals of
the present work. We notice here, however, that the running of the neutrino masses
is valid also for smaller values of the compact radii, even if in this case the effect
of the running becomes much smaller and it will be harder to detect experimentally.   

%%%%%%%%%%%%%%%%%%%%%%%%%%%%%%%%%%%%%%%%%%%%%%%%%%%%%%%%%%%%

%%%%%%%%%%%%%%%%%%%%%%%%%%%
\section*{Appendix A: Jacobi function}
%%%%%%%%%%%%%%%%%%%%%%%%%%%

The Jacobi function $\theta_3$ used in the text is defined as 
\begin{equation}
	\label{theta3two} 
\theta_3(z|\tau) = \sum_{k= - \infty}^{\infty}
e^{ 2 \pi i k z + i \pi k^2 \tau} \ . 
\end{equation}
A useful modular
transformation property is 
\begin{equation}
\theta_3 \left({z \over \tau}, {-1
\over \tau}\right)= \sqrt{-i \tau} \ e^{{i \pi z^2 / \tau}} \
\theta_3 (z, \tau ) \ . 
	\label{a4} 
\end{equation}
The asymptotic limits of
$\theta_3$ that we used are 
\be 
\lim_{t \rightarrow  \infty} \
\theta_3(z|\tau = i t) = 1 \quad , \quad \lim_{t \rightarrow  0} \
\theta_3(z|\tau = i t) \sim {1 \over
  \sqrt{t}} \ e^{- {\pi z^2 \over t}} \ .
\ee
We denote $\theta_3 (0|\tau)$ by $\theta_3 (\tau)$.

We can rewrite the eigenvalue equation~(\ref{eigenvalue}) by using the Jacobi~$\theta_3$ function. Indeed, by introducing the Schwinger proper time parameter $t$ through
\be
{1 \over A} \ = \ \int_0^{\infty} dt \ e^{- t A} \  \
\ee
and using the analytic continuation prescription $m = i \mu$, we find
\be
\sum_{k_1,k_2 = -\infty}^{\infty}
\frac{1}{m^2 -k_1^2/R_1^2 - k_2^2/R_2^2 } =  -
\int_{1/\Lambda^2}^{\infty}  \ d t \
 e^{- \mu^2 t} \ \theta_3 \left( \frac{i t}{\pi R_1^2} \right) \theta_3
 \left( \frac{i t}{\pi R_2^2} \right) \ .
\ee

Let us also mention the identity
\begin{equation}
	\label{eq:identitysum}
\sum_{(k_1,k_2) \in \mathcal{I}} \frac{2}{2^{\delta_{k_1 0} \delta_{k_2 0}}} f(k_1,k_2)
= \sum_{k_1,k_2=-\infty}^{\infty} f(k_1,k_2) \ , 
\end{equation}
valid for any function $f$ such that $f(-k_1,-k_2)=f(k_1,k_2)$ and where
$\mathcal{I}$ is the set defined in~eq.(\ref{eq:setI}).

%%%%%%%%%%%%%%%%%%%%%%%%%%%
\section*{Appendix B: 6D Dirac matrices}
%%%%%%%%%%%%%%%%%%%%%%%%%%%

For completeness, we give in this appendix the convention about spinors and Dirac matrices used
throughout the paper. We have mainly followed the conventions of Wess and Bagger~\cite{wb}.

In 6D, the Dirac matrices are $8\times 8$ and they can be easily constructed from the Dirac matrices in 4D:
\begin{equation}
\Gamma^\mu =
\left(
\begin{array}{cc}
\gamma^\mu & 0 \\
0 & \gamma^\mu
\end{array}
\right)_{\mu=0,1,2,3}\, ,
\
\Gamma^5 =
\left(
\begin{array}{cc}
0 & i \gamma^5 \\
i \gamma^5 & 0
\end{array}
\right)
\ \ \mathrm{and} \ \
\Gamma^6 =
\left(
\begin{array}{cc}
0 & - \gamma^5 \\
 \gamma^5 & 0
\end{array}
\right)
\end{equation}
with
\begin{equation}
\gamma^\mu = \left(
\begin{array}{cc}
0 & \sigma^\mu \\
\bar{\sigma}^\mu & 0
\end{array}
\right)
\ \ \mathrm{and} \ \
\gamma^5 = i \gamma^0 \ldots \gamma^3 =
\left(
\begin{array}{cc}
-\sigma^0 & 0 \\
0 & \sigma^0
\end{array}
\right)
\end{equation}
where $\sigma^\mu$ and $\bar{\sigma}^\mu$ are the usual Pauli matrices
\begin{eqnarray}
&
\sigma^0 = - \mathbf{1}_2, \,
\sigma^1 = \left( \begin{array}{cc}
0 & 1 \\
1 & 0
\end{array}
\right),
\sigma^2 = \left(  \begin{array}{cc}
0 & -i \\
i & 0
\end{array}
\right),
\sigma^3 = \left( \begin{array}{cc}
1 & 0 \\
0 & -1
\end{array}
\right),
\\
& \tv{13}
\bar{\sigma}^0 = \sigma^0, \,
\bar{\sigma}^1 = -\sigma^1, \,
\bar{\sigma}^2 = -\sigma^2, \,
\bar{\sigma}^3 = -\sigma^3.
\end{eqnarray}
A famous relation about the Pauli matrices that is useful to link the fermionic equation of motion to the scalar equation of motion is
\begin{equation}
\sigma^\mu \bar{\sigma}^\nu + \sigma^\nu \bar{\sigma}^\mu
= 2 \eta^{\mu \nu}
\ \ \mathrm{and} \ \
\bar{\sigma}^\mu \sigma^\nu + \bar{\sigma}^\nu \sigma^\mu
= 2 \eta^{\mu \nu}.
\end{equation}

The chirality matrix in 6D is simply defined by
\begin{equation}
	\label{eq:6DWeyl}
\Gamma^7 = \Gamma^0\ldots \Gamma^3 \Gamma^5 \Gamma^6 =
\left(
\begin{array}{cc}
\gamma^5 & 0 \\
0 & -\gamma^5
\end{array}
\right)\ .
\end{equation}
Therefore a 6D Weyl fermion is written in terms of a pair of  four-component Weyl fermions
\begin{equation}
\Psi =
\left(
\begin{array}{c}
\lambda_1 \\
\lambda_2
\end{array}
\right)\ ,
\end{equation}
with
\begin{equation}
\Psi = - \Gamma^7 \Psi \ \ \Leftrightarrow \ \
\lambda_1 = - \gamma^5 \lambda_1,
\
\lambda_2 =  \gamma^5 \lambda_2 \ . \label{weyl}
\end{equation}
The four-component Weyl fermions can be themselves conveniently written as two-components spinors
(we'll use the same notation for a four and two component spinor, the distinction should be clear from the context)
\begin{equation}
\lambda_1 =
 \left( \begin{array}{c} 0 \\ \bar{\lambda}^{\dot{\alpha}}_1 \end{array} \right) \ , \
 ~~~~
\lambda_2 =
\left( \begin{array}{c} \lambda_{2\, \alpha} \\ 0 \end{array} \right) \ .
\end{equation}

The dotted and undotted indices of a two-component spinor are raised and lower
with the $2 \times 2$ antisymmetric tensors $\epsilon_{\alpha \beta} = i \sigma^2_{\alpha \beta}$
and $\epsilon_{\dot{\alpha} \dot{\beta}} = i \sigma^2_{\dot{\alpha} \dot{\beta}}$ and their inverse
$\epsilon^{\alpha \beta}=-i \sigma^2_{\alpha \beta}$,
$\epsilon^{\dot{\alpha} \dot{\beta}} = -i \sigma^2_{\dot{\alpha} \dot{\beta}}$ :
\begin{equation}
\lambda_2^\alpha = \epsilon^{\alpha \beta} \lambda_{2\, \beta}
\ \ \mathrm{and}\ \
\bar{\lambda}_{1\, \dot{\alpha}} = \epsilon_{\dot{\alpha} \dot{\beta}} 
\bar{\lambda}_1^{\dot{\beta}} \ .
\end{equation}
Note also the adjoint relation:
\begin{equation}
(\lambda_2^\dagger)^\alpha = \bar{\lambda}_2^{\dot{\alpha}} \ .
\end{equation}
Finally, $\lambda_1 \lambda_2$ and $\bar{\lambda}_1 \bar{\lambda}_2$ denote the two Lorentz invariant scalars:
\begin{equation}
\lambda_1 \lambda_2 = \lambda_1^\alpha \lambda_{2\, \alpha}
\ \ \mathrm{and}\ \
\bar{\lambda}_1 \bar{\lambda}_2 = \bar{\lambda}_{1\, \dot{\alpha}} 
\bar{\lambda}_2^{\dot{\alpha}} \ .
\end{equation}
These products are symmetric
\begin{equation}
\lambda_1 \lambda_2 = \lambda_2 \lambda_1
\ \ \mathrm{and}\ \
\bar{\lambda}_1 \bar{\lambda}_2 = \bar{\lambda}_2 \bar{\lambda}_1 \ .
\end{equation}

The charge conjugation matrix in six dimensions, satisfying 
\be C
\ \Gamma^M \ C^{-1} \ = \ -  (\Gamma^M)^T \ , 
\ee
can be expressed
in terms of the 4D one as
\begin{equation}
C  \ = \ \left(
\begin{array}{cc}
0 & C_4 \\
-C_4 & 0
\end{array}
\right) \ ,
\end{equation}
where 
\be 
C_4 \ \gamma^{\mu} \ C_4^{-1} \ = \ -  (\gamma^{\mu})^T
\ . \ee
In the chiral representation of the Dirac matrices that we are using,
\begin{equation}
C_4 = i \gamma^0 \gamma^2 = \left( \begin{array}{cc} i\sigma^2 & 0 \\ 0 & - i \sigma^2 \end{array} \right)
\end{equation}
It can be easily checked that
\begin{equation}
C \ \Gamma^7 = - \Gamma^7 \ C
\ \ {\rm and}
\ \
C^{\dagger} \ = \ C\ .
\end{equation}
The charged conjugated spinor $\Psi^c$ is defined through
\be
\Psi^c = C \bar{\Psi}^T \ .
\ee
It can be easily checked that for the 6D Weyl spinor satisfying eqs.~(\ref{weyl}), 
we have (in four and two component notations):
\begin{eqnarray}
& \displaystyle
  {\bar \Psi}  \Psi \ = \ 0 \ ,  \ {\bar \Psi}^c  \Psi \ = \ 0 \ ,
\\
& \displaystyle
{\bar \Psi}^c  \Gamma^5 \Psi =
 i \left( \lambda_2^t C_4 \lambda_2 +  \lambda_1^t C_4 \lambda_1 \right)
= - \left( \bar{\lambda}_1 \bar{\lambda}_1 + \lambda_2 \lambda_2 \right) \ ,
\\
& \displaystyle
{\bar \Psi}^c \Gamma^6 \Psi =
 - \lambda_2^t C_4 \lambda_2 +  \lambda_1^t C_4 \lambda_1
= -  \bar{\lambda}_1 \bar{\lambda}_1 + \lambda_2 \lambda_2  \ .
\end{eqnarray}

\vspace{1cm}

%%%%%%%%%%%%%%%%%%%%%%%%%%%
\mysection{Acknowledgements}: We are grateful to G.~Bhattacharyya, T. Gherghetta,
P.~Holstein, S.~Lavignac, J.~Mourad, C.~Papineau, M. Peloso, K.~Sridhar  and especially
to V.~Rubakov for suggestions and discussions. SKV acknowledges
support from Indo-French Centre for Promotion of Advanced Research
(CEFIPRA) project No:  2904-2 `Brane World Phenomenology'. We all acknowledge
the hospitality of the Department of Theoretical Physics of TIFR-Mumbai, whereas
E.D. would like to thanks the hospitality of the William I. Fine Theoretical 
Physics Institute of the University of Minnesota while completing this work. 
Work partially supported by INTAS grant, 03-51-6346, CNRS PICS 
\#~2530 and 3059, RTN contracts MRTN-CT-2004-005104 and MRTN-CT-2004-503369, a European 
Union Excellence Grant, MEXT-CT-2003-509661 and by the ACI Jeunes Chercheurs 2068.

%%%%%%%%%%%%%%%%%%%%%%%%%%%

\end{document}